# An Approach to Ensure Fairness in News Articles

Shaina Raza
University of Toronto
Toronto, Canada
Shaina.raza@utoronto.ca

Deepak John, Reji
Environmental Resources Management
Bangalore, India
deepak.reji@erm.com

Dora D. Liu
DeepBlue Academy of Sciences
China
liudongmei_0506@163.com

Syed Raza, Bashir
Toronto Metropolitan University
Toronto, Canada
syedraza.bashir@ryerson.ca

Usman Naseem
The University of Sydney,
Sydney, Australia
usman.naseem@sydney.edu.au

## ABSTRACT

Recommender systems, information retrieval, and other information access systems present unique challenges for examining and applying concepts of fairness and bias mitigation in unstructured text. This paper introduces Dbias (https://pypi.org/project/Dbias/), which is a Python package to ensure fairness in news articles. Dbias is a trained Machine Learning (ML) pipeline that can take a text (e.g., a paragraph or news story) and detects if the text is biased or not. Then, it detects the biased words in the text, masks them, and recommends a set of sentences with new words that are bias-free or at least less biased. We incorporate the elements of data science best practices to ensure that this pipeline is reproducible and usable. We show in experiments that this pipeline can be effective for mitigating biases and outperforms the common neural network architectures in ensuring fairness in the news articles.

## 1. INTRODUCTION
Information retrieval, recommendation systems and information access systems are often trained on a large text corpus which may introduce biases into the models. These biases can arise in a variety of contexts, including observations that comprise the data, training, development, evaluation, and application of the underlying models [21]. Data-centric systems, like news recommender systems, are also trained on massive amounts of data with little or no control over the quality of training data [25].

Research [18] shows that it is highly important to eliminate these biases early in the data gathering process, before they enter the system and are reinforced by model predictions, resulting in biases in the model decisions. [7,17].

Bias and fairness are hot topics both in academia and industry. Recently, some comprehensive surveys have emerged on these topics, shedding light on the source(s) of bias and potential solutions [17,21,30]. Bias can be, conventionally, defined as an anomaly in the data or output of an ML algorithm caused by prejudiced assumptions [17], such as related to gender, race, demographics, economic status, or religion [17]. Bias in natural language processing (NLP) [5], generally, refers to harmful prejudices against certain groups expressed as toxic or offensive words [2,5,11]. The goal of fairness is to identify and mitigate the effects of various biases [20], as well as to build ML models to prevent repeating human and societal biases or adding new biases. In this work, we aim to eliminate biases in the text data.

According to research [2,24,27], news media can be extremely biased, and biased news can result in "filter bubbles" or "echo chambers" [6,25], which may lead to a lack of understanding of specific issues and a narrow, one-sided perspective. This motivates us to train Dbias to counteract media bias. We summarize our contributions as:

(1) We develop a fair ML pipeline, which we name as Dbias (de-biasing). This pipeline consists of multiple sequential steps that perform bias detection, bias recognition, and bias mitigation tasks, as well as recommend bias-free information.

(2) We make Dbias available as a python package that includes documentation, usage, and tutorials to assist data scientists and practitioners in integrating this package into their work products.

(3) Dbias is released as a reusable and self-contained pipeline that has its algorithms for detecting and mitigating biases. This makes it different from existing fair ML pipelines [1,3,4,28], which use off-the-shelf bias detection or bias mitigation models. We develop this pipeline according to the widely accepted data science pipeline structure [12], so this pipeline is reusable. The only requisite is to train the model on domain-specific data.

The rest of the paper is organized as: Section 2 is the proposed approach. Section 3 is experiments and results. Section 4 is the conclusion.

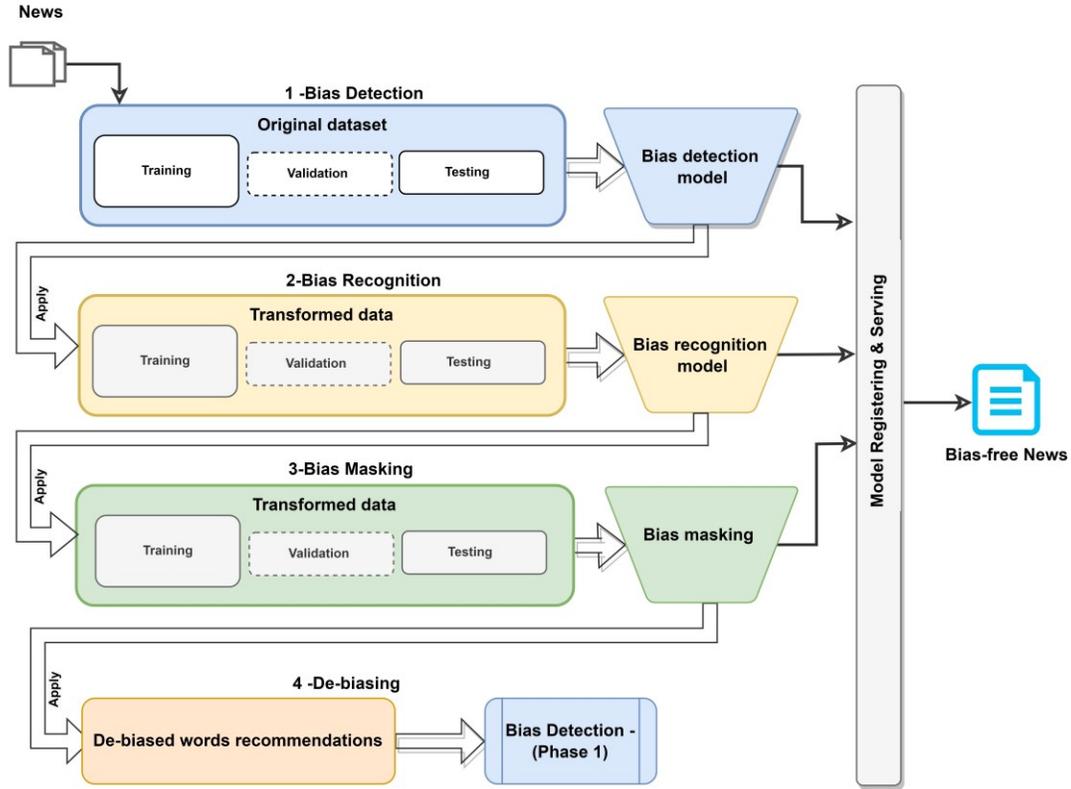

Figure 1: Dbias, a fair ML pipeline and its workflow

## 2. PROPOSED APPROACH

### 2.1 Problem Definition

Given a collection of news articles that may contain a variety of biases, the objective is to detect, recognize, and eliminate these biases from the data. The goal is to enable end-users to shift seamlessly from raw data that may be biased to a fair model. Next, we explain each phase of the Dbias pipeline.

### 2.2 Data

In this work, we primarily use MBIC – A Media Bias Annotation Dataset [27], which consists of news articles from different news sources (HuffPost, MSNBC, USA Today and others). We use the extended version of MBIC dataset[1] containing thousands of records, with both biased and non-biased sentences. In the original dataset, these biases are identified through crowdsourcing. We identified more biases (these biases are related to gender, race, ethnicity, education, religion, language) from the literature [11,15] and also added some biases manually. The dataset features used in this work are news snippet; URL; news source; topic; age; gender; education; biased words and label.

### 2.3 Methodology

The specific tasks to perform in this work are:
*Bias detection*: To detect whether a news article is biased or not.
*Bias recognition:* To recognize the biased words/ phrases in news.
*Bias masking:* To mask (hide) the biased words.
*De-biasing:* To de-bias the data by replacing the biased words with the unbiased or less biased word(s).

---
[1] https://zenodo.org/record/5861846#.YoT6wajMK5d

#### 2.3.1 Bias Detection

This is the first module in the Dbias pipeline, as shown in Figure 2. We use the DistilBert (a distilled version of BERT) and fine-tune it on the MBIC dataset. For binary classification, we use binary-cross entropy loss [19] with a sigmoid activation function. The output from the bias detection model is a set of news articles that are classified as biased or non-biased.

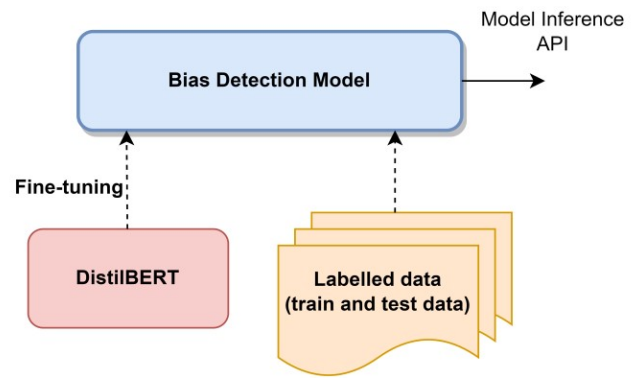

Figure 2: Bias detection module

#### 2.3.2 Bias Recognition

The second module in the Dbias pipeline is the bias recognition module. This task is different from the bias detection task that takes a whole news or sentence and classifies if it is biased or not. In bias recognition, we use the named entity recognition task of NLP to identify the biased words from the text.

This module takes as input a set of news articles that have been identified as biased in the preceding module (bias detection), and outputs a set of news articles with biased words that are identified and recognized. For example, the news "Don't buy the pseudo-scientific hype about tornadoes and climate change" has been classified as biased by the preceding bias detection module, and the biased recognition module can now identify the term "pseudo-scientific hype" as a biased word.

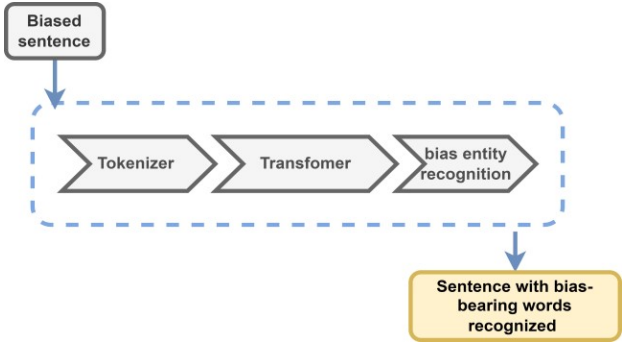

**Figure 3: Bias recognition module**

Though, the standard named entity recognition (NER) task is not directly related to bias identification, it can be used to find biases in data [9,17]. For example, one NER model [16] has been used to find if there are more female names tagged as non-person than male names. Another NER model has identified biases based on occupation, race, and demographics [9]. In this work, we refer to each entity in the NER as a bias-bearing entity that is manifested in syntax, semantics or linguistic context.

We use the RoBERTa [14] along with Spacy English Transformer NER pipeline [10]. The final output from the bias recognition module is a set of news articles, where the biased words have been identified. We show our bias recognition module, which is a pipeline, in Figure 3.

### 2.3.3 Bias Masking

*Bias Masking:* In the bias masking stage, we mask the position of each biased word (token) within each news article. We use Masked Language Modeling (MLM) [26] for masking biased words. Typically, the MLM task takes a sentence, randomly masks 15% of words in the input and then runs the entire masked sentence through the model to predict masked words.

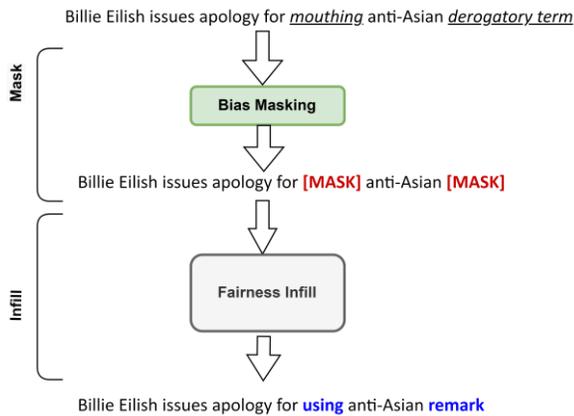

**Figure 4: Bias masking example**

In our work, we only mask those words that have been flagged as biased by the previous bias recognition module. We propose a unique mask shifting technique that can mask and unmask more than one token at a time in a sentence. For example, as shown in Figure 4, we see that both instances are processed sequentially via the mask shifting technique, and each masked token is filled one at a time before the final sentence is constructed.

*Fairness infilling*: we propose a fairness infilling stage, which can be considered as generalizing the cloze task (Wu et al. 2019) from single tokens to spans of unknown length. Our assumption behind this fairness filling is that the new words that are filled in are less or non-biased, which has been validated through our demonstration and experiments.

### 2.3.4 De-biasing and recommendations

We recommend a couple (5, 10, 15 or so) of substitute tokens that can be used to infill each masked token during the fairness infilling stage. We send the top-k recommended sentences (infilled with new tokens) again to the bias detection model (first module) to see the probability of biasness. If the probability of biasness is less than 0.5 or less than the probability of the previous de-biased sentence, we output the sentence as the final output.

## 3. EXPERIMENTS AND RESULTS

### 3.1 Experimental Setup

We implemented this pipeline in TensorFlow. We run our experiments on Google Colab Pro (NVIDIA P100, 24 GB RAM, 2 x vCPU). Common parameters used in Dbias are a batch size of 16, 10 epochs, sequence length 512, and number of labels for news is 2 (bias, non-biased). We use the distilbert-base-uncased with these details: Uncased: 6-layer, 768-hidden, 12-heads, 66M parameters, in the bias detection module. We use RoBERTa-base, 12-layer, 768-hidden, 12-heads, 125M parameters in the bias recognition module. The learning rate, warm-up setups, the drop-out rate and other parameters for each module are optimized according to their best settings. For a fair comparison, we tune all the other methods (our method and baselines) to their optimal hyperparameter settings and report the best results.

To assess the performance of our proposed model, we use the accuracy (ACC), precision (PREC), recall (Rec) and F1-score (F1), following standard metrics in this line of research [4]. To our knowledge, there is no single model or pipeline that can perform all three tasks simultaneously, so, we evaluate Dbias performance for main tasks (bias detection and recognition).

### 3.2 Results and Analysis

The results are shown and discussed next:

#### 3.2.1 Effectiveness of bias detection module

In this experiment, we evaluate the performance of our framework for the bias detection module (the first module of the Dbias) against the following state-of-the-art baselines:

*LG-TFIDF*: We use the Logistic Regression (LG) [29] with TfidfVectorizer (TFIDF) word embeddings[22].
*LG-ELMO*: We use LG with ELMO embeddings [23].
*MLP-ELMO:* We use MultiLayer Perceptron (MLP), a feedforward artificial neural network with ELMO embeddings.
*BERT*: We use Bidirectional Encoder Representations from Transformers (BERT) [8] and its bert-based uncased version.

*RoBERTa*: We use Robustly Optimized BERT Pre-training Approach (RoBERTa) [13] with RoBERTa-base.
The results are shown in Table 1.

**Table 1: Performance of bias detection task**

| Model | PREC | REC | F1 |
|---|---|---|---|
| LG-TFIDF | 62% | 61% | 62% |
| LG- ELMO | 67% | 69% | 68% |
| MLP- ELMO | 69% | 68% | 68% |
| Bert | 72% | 69% | 71% |
| RoBERTa | 76% | 70% | 73% |
| **Our approach** | **77%** | **74%** | **75%** |

Overall, the results in Table 1 show the better performance of our approach (based on distilBERT fine-tuned on MBIC dataset) compared to the baseline methods for the bias detection task. The result also shows that deep neural embeddings, in general, can outperform traditional embedding methods (e.g., TFIDF) in the bias classification task. This is shown by the better performance of deep neural network embeddings (i.e., ELMo) compared to TFIDF vectorization when used with LG. This is probably because deep neural embeddings can better capture the context of the words in the text in different contexts. The deep neural embeddings and deep neural methods (MLP, BERT, RoBERTa) also perform better than traditional ML method (LG).

We also observe that Transformer-based methods outperform other methods (ML and simple deep learning methods) in the bias detection task. Among the Transformer-based approaches, RoBERTa outperforms the BERT model by approximately 2% in the F1-score, while our approach based on DistilBERT outperforms the RobBERTa by ~2%. DistilBERT is smaller, faster, and lighter than BERT and RoBERTa. When we use DistilBERT in our work, it also performs better than all the other models. So, we choose to work with the DistilBERT for the bias detection task.

### 3.2.2 Effectiveness of bias recognition module
We choose the following baselines for NER task based on similar structure as in our bias recognition module.
Spacy core web small pipeline (core-sm)[2].
Spacy core web medium pipeline (core-md)[3]
Spacy core web large pipeline (core-lg)[4]
Our approach is based on *Spacy core web transformer pipeline* (core-trf)[5]. The results of different NER methods are shown in Table 2.

**Table 2: Performance of bias recognition task**

| Model | PREC | REC | F1 | ACC |
|---|---|---|---|---|
| **Core-sm** | 59% | 27% | 37% | 37% |
| **Core-md** | 61% | 45% | 52% | 53% |
| **Core-lg** | 60% | 62% | 60% | 67% |
| **Our approach** | **66%** | **65%** | **63%** | **72%** |

The results in Table 2 show that our approach based on core-trf outperforms all the other NER methods in terms of precision-recall, F1-score and accuracy. This is because a Transformer-based model can identify entities and relations within the text with rich contexts. We also find that bias recognition tasks show better accuracy with larger model size. This is likely because the larger model contains a greater number of parameter settings and data points, all of which affect the model's predictive performance. However, these benefits come at the expense of resource utilization, memory, CPU cycles, and latency delay. Based on these results, we choose to work with core-trf.

### 3.3 Working of Dbias
We release Dbias as a python package that can be used to detect and mitigate biases in texts. The input to the model can be any sentences that may contain biased words and we get the de-biased output. We show a working example based on a piece of biased news in Figure 5.

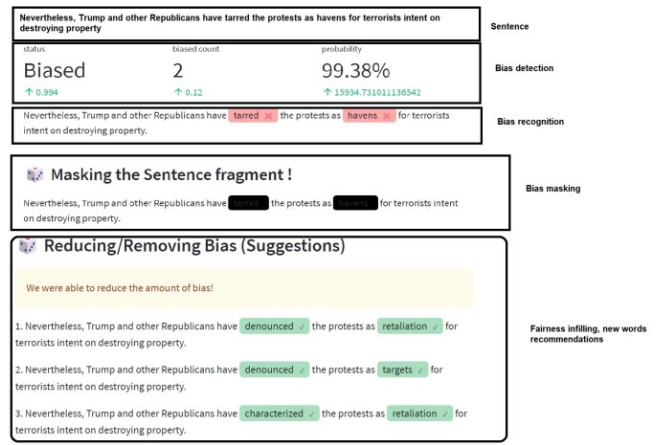

**Figure 5: Example on Dbias**

As illustrated in Figure 5, given a news article or any text that may contain biased words, our model can determine whether or not the text is biased. This is made possible by the first module: the bias detection module. The output is then forwarded to the next module, namely the bias recognition module, which identifies bias-bearing words. The text with identified biased words is then sent to the de-biasing module, which masks the biased words and makes suggestions for new words to replace them. The final output is a set of non-biased or at least minimally biased sentences for each input sentence.

## 4. Conclusion
We build Dbias, a pipeline for fair ML that has stages: bias detection, bias recognition, bias masking and de-biasing. We develop Dbias as a downloadable package that can be used as it or integrated in a system like a news recommender system or so. This research serves as a forum for researchers interested in de-biasing the text.

*Limitations and future works:* We need to investigate a wide variety of biases in news media to determine the definitions of biases and that of fairness. So far, we use a manually annotated news dataset to identify bias-bearing words. We encourage researchers to identify more biases in the text to annotate one such data. The fundamental requirement for our system is to fine-tune it using the domain data. While this is a start, we may need to investigate additional domains to enrich the Dbias pipeline.

---
[2] en_core_web_sm
[3] en_core_web_md
[4] en_core_web_lg
[5] en_core_web_trf